%% file: screl_arxiv.tex
\documentclass{article}

\usepackage{import} %

\import{includes/}{newcommands}
\usepackage{amsmath}
\usepackage{amssymb}
\usepackage{graphicx}
\usepackage{epstopdf, epsfig}
\usepackage{xcolor}

\newcommand{\upi}{\pi}

\title{Inverse transfer of magnetic helicity in direct numerical simulations of  compressible isothermal turbulence: scaling laws}

\author{Jean-Mathieu Teissier$^1$ and Wolf-Christian M\"uller$^{1,2}$}

\date{\small $^1$ Technische  Universit\"{a}t Berlin, ER 3-2, Hardenbergstr. 36a, D-10623 Berlin, Germany \\
$^2$ Max-Planck/Princeton Center for Plasma Physics\\ \vspace{1em}
\today}

\usepackage{siunitx} %
\definecolor{corrclr}{rgb}{0,0,0}
\definecolor{corrclrT}{rgb}{0,0,0}
\definecolor{corrclrW}{rgb}{0,0.6,0}

\usepackage{amsmath}

\newcommand{\grpath}{figures/}
\graphicspath{{\grpath}}

\def\plotvspaceval{0}
\def\plotvspacevalend{0}
\def\plotvspacemid{0}

\begin{document}

\maketitle

\begin{abstract}

The inverse transfer of magnetic helicity is investigated through direct numerical simulations of large-scale-mechanically-driven turbulent flows in the isothermal ideal magnetohydrodynamics (MHD) framework. The mechanical forcing is either purely solenoidal or purely compressive and the turbulent steady-states considered exhibit root mean square (RMS) Mach numbers $0.1 \lesssim \rmsM \lesssim 11$. A continuous small-scale electromotive forcing injects magnetic helical fluctuations, which lead to the build-up of ever larger magnetic structures. Spectral scaling exponents are observed which, for low Mach numbers, are consistent with previous research done in the incompressible case. Higher compressibility leads to flatter magnetic helicity scaling exponents. The deviations from the incompressible case are comparatively small for solenoidally-driven turbulence, even at high Mach numbers, as compared to those for compressively-driven turbulence, where strong deviations are already visible at relatively mild RMS Mach numbers $\rmsM\gtrsim 3$. Compressible effects can thus play an important role in the inverse transfer of magnetic helicity, especially when the turbulence drivers are rather compressive. Theoretical results observed in the incompressible case can, however, be transferred to supersonic turbulence by an appropriate change of variables, using the Alfv\'{e}n velocity in place of the magnetic field.
\end{abstract}

\section{Introduction}

Magnetic helicity $\mhel=\volavg{\FmagA \cdot \Fmag}$, defined as the helicity of the magnetic vector potential $\FmagA$ (so that $\Fmag=\nabla \times \FmagA$ is the magnetic field), with $\langle \cdot \rangle$ denoting the volume average, is a quantity expressing topological aspects of the magnetic field lines, such as their degree of linkage, twist, writhe and knottedness \cite{MOF69}. In a closed volume with a vanishing magnetic field component normal to the volume boundaries, or a periodic domain without mean magnetic field, magnetic helicity is gauge invariant \cite{BIS93,BER97}. In space, because of the very low resistivity, magnetic field lines are effectively ``frozen-in'' the ionised gas (plasma) present in various degrees of ionisation throughout the universe \cite{ALF42}. As a consequence, rotational motion is expected to naturally generate helical magnetic fields. Since magnetic helicity is an ideal invariant of the magnetohydrodynamic (MHD) equations \cite{ELS56,WOL58a}, it can be an important constraint on the time evolution of magnetic fields, which plays a crucial role in many astrophysical systems of interest. For example, magnetic helicity dynamics are involved in solar flares and coronal mass ejections \cite{LOW94}, which transfer magnetic helicity from the sun to the interplanetary medium \cite{KUR96}. The sun also emits magnetic helicity through the solar wind \cite{BEM87}, which is manifested at the largest scales through the Parker spiral \cite{BER99}. Magnetic helicity conservation is also very important in dynamo processes \cite{VIC01,BRA01,BRL13}. In laboratory plasmas, it is of relevance for plasma confinement in reversed-field-pinch fusion experiments \cite{EMO00}.

Magnetic helicity is subject to an inverse transfer in spectral space, as has been suggested in \cite{FPL75} and subsequently confirmed by various numerical experiments \cite{PFL76,POP78,MFP81,BAP99,CHB01,BRA01,AMP06,MAL09,MMB12,LBM16,LSM17}. This key property renders magnetic helicity a potentially important quantity in the generation and sustainment of large-scale magnetic fields in the universe.

Up to the present day, the inverse transfer of magnetic helicity in single-fluid MHD has been studied assuming either an incompressible plasma \cite{AMP06,MAL09,MMB12}, or in the subsonic/transonic case \cite{BAP99,CHB01,BRA01}. In astrophysical systems however, flows are often highly supersonic. For example, the order of magnitude of the root mean square (RMS) turbulent Mach number ranges from 0.1 to about 10 in the interstellar medium (\cite{ELS04}, section 4.2). The present work makes hence one step towards a more realistic setting by including compressible effects, in the framework of supersonic isothermal ideal MHD. The inverse transfer of magnetic helicity is investigated through direct numerical simulations of large-scale-mechanically-driven compressible plasma under continuous injection of small scale random helical magnetic fluctuations. The large scale mechanical forcing is either purely solenoidal or purely compressive and the considered RMS Mach numbers of the initial hydrodynamic turbulent steady-state range from 0.1 to 11 for solenoidal driving and from 1 to 8 for compressive driving.

A similar numerical setup in the incompressible case, where magnetic helical fluctuations are injected at small scales in the absence of a large-scale mechanical forcing has shown that several quantities, including the kinetic and magnetic energies and helicities, exhibit approximate power-law scaling in Fourier space in the inverse transfer region, i.e. at spatial scales larger than those of the magnetic driving \cite{MAL09,MMB12}. Furthermore, a dynamic balance leading to a quasi-equipartition between magnetic and kinetic energies on the one hand and current and kinetic helicities on the other hand has been observed in several numerical experiments in the incompressible case \cite{MAL09,MIP09,GMP11,MMB12}. The purpose of the present work is to investigate similarities and differences in the spectral scaling laws and this dynamic balance for turbulent flows exhibiting significant compressible effects. The scaling properties are expected to change, since in compressible MHD, the magnetic field time evolution is governed by:

\beq
	\pat \Fmag = -(\Fvel \cdot \nabla) \Fmag+(\Fmag \cdot \nabla) \Fvel - \Fmag (\nabla \cdot \Fvel).
\eeq

For a high level of compression, the magnetic field and hence magnetic helicity dynamics are expected to be affected through two aspects. First, through the compression term $-\Fmag (\nabla \cdot \Fvel)$, which is not present in incompressible MHD. Second, even though the advective $-(\Fvel \cdot \nabla)\Fmag$ and the stretching $(\Fmag \cdot \nabla)\Fvel$ terms are also present in the incompressible case, the velocity field's compressive part may alter them. An analysis of the influence of the compressive velocity field on these three terms is the object of a forthcoming work.

The numerical method used as well as the simulation setup are described in section \ref{sec:numexp}. The Fourier space formalism is presented in section \ref{sec:notations}. Section \ref{sec:prevresearch} summarises some findings of previous research, which are put in relation with this work's results in section \ref{sec:results}. Finally, section \ref{sec:confirmhighres} assesses the robustness of the results while concluding remarks are given in section \ref{sec:conclusion}.

\section{Numerical experiments}
\label{sec:numexp}

\subsection{MHD numerical solver}

The isothermal ideal (single fluid) MHD equations, in the presence of both a mechanical and an electromotive forcing, can be written as:

\beqa
\pat \rho &=& - \nabla \cdot (\rho \Fvel),\\
\pat (\rho \Fvel) &=& - \nabla \cdot \left( \rho \Fvel \Fvel^T + (\rho \cs^2 + \frac{1}{2}|\Fmag|^2)\mathI - \Fmag \Fmag^T \right)+\rho\ForceV,\\
\pat \Fmag &=& \nabla \times (\Fvel \times \Fmag)+\ForceM,\\
\nabla \cdot \Fmag&=&0,
\eeqa

with $\rho$ the mass density, $\Fvel$ the fluid velocity, $\cs$ the constant isothermal sound speed so that $\press=\rho \cs^2$ is the (thermal) pressure and $\Fmag$ the magnetic field so that $\vE=-\Fvel \times \Fmag$ is the electric field. The $3\times 3$ identity matrix is denoted by $\mathI$. The mechanical driving occurs through an acceleration field ($\ForceV$) and the magnetic helical fluctuations are injected through an electromotive forcing ($\ForceM$). These two terms are described in section \ref{sec:forcingterms}.

The numerical solver is a shock-capturing fourth-order finite volume method, described in detail in \cite{TEI20,VTH19}. The main reconstruction method used is a fourth-order Central Weighted Essentially Non-Oscillatory (CWENO) procedure \cite{LPR99}, with a passage through point values in order to maintain the fourth-order accuracy \cite{COC11,BUH14}. The interfacial fluxes are computed using the Rusanov approximation \cite{RUS61}, also known as ``Local Lax-Friedrichs''. In order to prevent the appearance of negative densities in high Mach number flows, a local reduction of the reconstruction order in the vicinity of strong discontinuities and shocks is used (an approach often referred to as ``flattening'' or ``fallback approach''). This has the effect of adding some numerical diffusivity locally, which smoothens the solution. The magnetic field solenoidality is maintained up to machine precision by application of the constrained-transport approach \cite{EVH88}, making use of a multidimensional version of the Rusanov flux to compute the line-integrated electric field \cite{BAL10}. The time-integration is implemented as a fourth-order ten-stages Strong Stability Preserving Runge-Kutta (SSPRK) method (\cite{KET08}, pseudocode 3). The discrete timestep $\Dt$ is constrained by the Courant-Friedrichs-Lewy criterion with a Courant number of $C_{CFL}=1.5$.

\subsection{Simulation procedure}
The considered turbulent systems exhibit both a direct cascade of kinetic energy from large to small scales and an inverse transfer of magnetic helicity, from small to large scales. The terminology ``cascade'' is not used here for the magnetic helicity inverse transfer, since this process displays pronounced spectrally non-local features \cite{AMP06,TEI20}.
\label{sec:forcingterms}
The turbulent steady-state is generated and sustained through a continuous large scale mechanical driving in a triply periodic cubic simulation domain with linear size $\Lbox=1$. The isothermal sound-speed is taken as $\cs=0.1$. Starting with a constant-density plasma at rest ($\rho=\rhoz=1,\Fvel=0,\Fmag=0$), an Ornstein-Uhlenbeck driving injects kinetic energy at large scales, similarly to \cite{FRD10,FED13}. It is defined, starting with $\fOUFk=0$, by the following stochastic differential equation in Fourier space:

        \beq
                \label{eq:dtfOUF}
                d\fOUFk(t)=-\fOUFk(t)\frac{dt}{\fat}+\famp \left( \frac{2\sigma(\vk)^2}{\fat} \right)^{1/2} \projkT \cdot d\vW(t),
        \eeq

	with $\fat=\Lbox/(2\cs\rmsM^{*})$ the forcing auto-correlation time, where $\rmsM^*$ is an a priori estimate of the expected RMS Mach number at steady-state, $\famp$ an amplitude, $\sigma$ a spectral profile, $d\vW(t)=dt\vN(0,dt)$ a Wiener process with $\vN(0,dt)$ a 3D Gaussian distribution with zero mean and standard deviation $dt$ representing a three-dimensional continuous random walk. The projection operator $\projkij(\vk)=\specw \delta_{ij}+(1-2\specw) \frac{k_ik_j}{|k|^2}$ allows to control the forcing's compressivity: for $\specw=0$, only components along $\vk$ are kept so that the forcing is purely compressive, whereas for $\specw=1$, the forcing is projected on the plane orthogonal to $\vk$ in Fourier space and the resulting field is purely solenoidal. The spectral profile is taken as $\sigma(\vk)=1$ for the wavenumber shells $1 \leq \kspec \leq 2$ and 0 otherwise so that only the largest scales are forced. The wavenumber shell $\kspec \in \mathbb{N}$ is defined by the wavevectors $\vk$ such that $\kspec \leq |\vk|/\kone < \kspec+1$ with $\kone=2\upi/\Lbox$ the smallest wavenumber in the system. The forcing is applied after the ten stages of the SSPRK integration method, through:

\beq
      (\rho \Fvel) \leftarrow (\rho \Fvel)+\rho \Dt \ForceV, %
\eeq

	with $\ForceV=\AmpV \tForceV$, where $\tForceV$ is the $\fOUF$ field in configuration space, multiplied by an amplitude $\AmpV$. This amplitude is introduced in order to guarantee a constant energy injection rate $\Einj=\Delta \Ekin/\Dt$ and is determined by the largest root of the second-order equation (cf. \cite{MCL99}):

\beq
        \label{eq:2ndorder_ekin}
        \Delta \Ekin =\frac{1}{2} \AmpV^2 \Dt^2 \sum_{i,j,k} \rho_{i,j,k} \tForceVijk^2+\AmpV\Dt \sum_{i,j,k} \rho_{i,j,k} \Fvel_{i,j,k} \cdot \tForceVijk,
\eeq

	where $q_{i,j,k}$ designates the value of field $q$ in the numerical cell indexed by $(i,j,k)$. As a consequence of this normalisation, the choice of the constant amplitude $\famp$ in relation \eqref{eq:dtfOUF} is arbitrary. The injected energy cascades to ever smaller scales where it is finally dissipated due to numerical non-ideal effects, which serve as a simple model for physical viscosity and resistivity. Thus, a steady-state is reached with a roughly constant RMS Mach number $\rmsM = \sqrt{ \volavg{|\Fvel|^2/\cs^2} } \approx \rmsM^*$ on a time scale of the order of the turbulent turnover-time $\ttt=\Lbox/(2\cs\rmsM)$ \cite{FED13}. The energy injection rate determines the steady-state RMS Mach number $\rmsM$. 
	The weak fluctuating mean velocity field which appears as a result of the driving is removed at each iteration.

	At a particular instant in time during the turbulent hydrodynamic statistical steady-state, a small scale electromotive helical driving is switched on while maintaining the large-scale mechanical driving. The electromotive forcing injects delta-correlated (white noise) maximally helical magnetic fluctuations at the wavenumber shells $48 \leq \kspec \leq 52$ and is defined in Fourier space by:

\beq
	\fMAGk=\Ampp e^{i\phasep}\vhhs{+}{\vk},
\eeq

	with $\phasep$ a random phase chosen uniformly in $[0,2\upi]$ and $\vhhs{+}{\vk}$ an eigenvector of the curl operator such that $i\vk \times \vhhs{+}{\vk}=+|\vk| \vhhs{+}{\vk}$ (see e.g. \cite{BRS05}). The amplitude $\Ampp$ in the different shells is given by:

\beq
	\Ampp=\exp\left[\frac{1}{2}\left(\frac{|\vk|/\kone-50}{4}\right)^2\right].
\eeq

	Similarly to the mechanical forcing, the resulting field in configuration space, $\tForceM$, is applied after the ten stages of the time integration procedure. It is renormalised in a way analogous to $\tForceV$ in order to guarantee a constant magnetic energy injection rate $\EMinj=\Delta \Emag/\Dt$.

\subsection{Simulation runs}
\label{sec:performedruns}

\begin{table}
  \begin{center}
\def~{\hphantom{0}}
  \begin{tabular}{cccccc}
      Label  & $\Einj$   & $\specw$  & $\fat$ & $\rmsM$ & $\EMinj$ \\[3pt]
\hline
       M01s4   & 7$\times 10^{-7}$ & 1 & 50 & 0.116 & 28$\times 10^{-7}$ \\
       M1s2   & 7$\times 10^{-4}$ & 1 & 5 & 1.09 & 14$\times 10^{-4}$ \\
       M5s  & 0.11 & 1 & 1 & 5.06 & 0.11 \\
       M7s   & 0.31 & 1 & 5/7 & 7.03 & 0.31  \\
       M11s & 1.21 & 1 & 5/12 & 11.1 & 1.21  \\
  \end{tabular}
  \begin{tabular}{ccccccc}
      Label  & $\Einj$   & $\specw$  & $\fat$ & $\rmsM$ & $\EMinj$ \\[3pt]
\hline
       M1c   & 7$\times 10^{-4}$ & 0 & 5 & 0.797 & 7$\times 10^{-4}$ \\
       M3c   & 1.9$\times 10^{-2}$ & 0 & 5/3 & 2.80 & 1.9$\times 10^{-2}$ \\
       M5c, M5cB   & 0.11 & 0 & 1 & 5.05 & 0.11 \\
       M8c   & 0.31 & 0 & 5/7 & 7.87 & 0.31 \\
          &  &  &  &  & \\
  \end{tabular}
  \caption{Parameters of the simulation: kinetic energy injection rate $\Einj$, spectral weight $\specw$ governing the forcing's compressivity, forcing auto-correlation time $\fat$, which result in a time-averaged root mean square Mach number $\rmsM$. During the hydrodynamic steady-state, magnetic helicity injection is switched on with a constant magnetic energy injection rate $\EMinj$.}
  \label{tab:runsparam}
  \end{center}
\end{table}
	The forcing parameters used for the simulated runs and the resulting time-averaged RMS Mach number in the hydrodynamic steady-state $\rmsM$ are displayed in table \ref{tab:runsparam}. The number in the runs' labels stands for their approximate $\rmsM$ and the letter (`s' or `c') for the forcing used (purely solenoidal with $\specw=1$ or purely compressive with $\specw=0$). The main runs are performed at resolution $512^3$ and confirmation at higher resolution has been done for selected cases, see section \ref{sec:confirmhighres}. In order to investigate subsonic, transonic and supersonic turbulence at various compressibility, $\rmsM$ varies in the range $\approx 0.1-11$ for the solenoidally-driven runs and $\approx 1-8$ for the compressively-driven ones. The magnetic-to-kinetic energy injection ratio $\EMinj/\Einj$ is taken as unity for all runs, apart for the M01s4 and M1s2 runs where it is 4 and 2 respectively (as written in the run's label) in order to obtain a faster convergence of the scaling laws in the inverse transfer region. A parameter study in order to assess the influence of the magnetic-to-kinetic energy injection ratio has been performed in \cite{TEI20}, section 7.1 and suggests that for $\EMinj$ and $\Einj$ of the same order of magnitude, the scaling exponents in the magnetic helicity inverse transfer region converge to the same value, this convergence being faster with a greater magnetic energy injection rate $\EMinj$.

\def\plotvspaceval{-6}
\def\plotvspacevalend{-1}
	 
 \begin{figure}[h]
     \centering
     \large
     \resizebox{1.\linewidth}{!}{\input{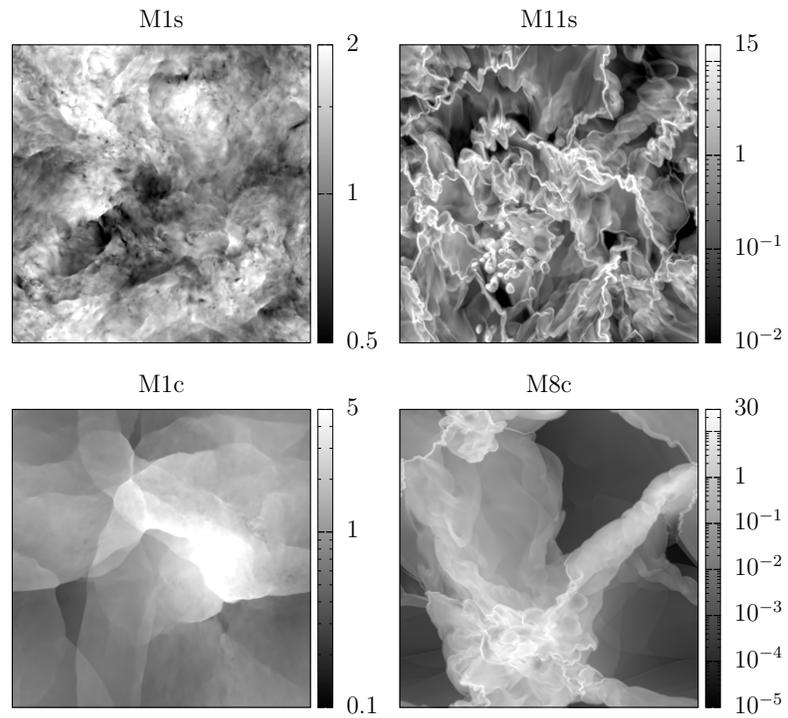}}	
     \caption[Small caption]{\footnotesize{Mass density slices during the hydrodynamic steady-state for the M1s2, M11s, M1c and M8c runs.}}
     \label{fig:Hdenslices}
 \end{figure}

\def\plotvspacevalend{0}
\def\plotvspaceval{0}

\begin{figure}
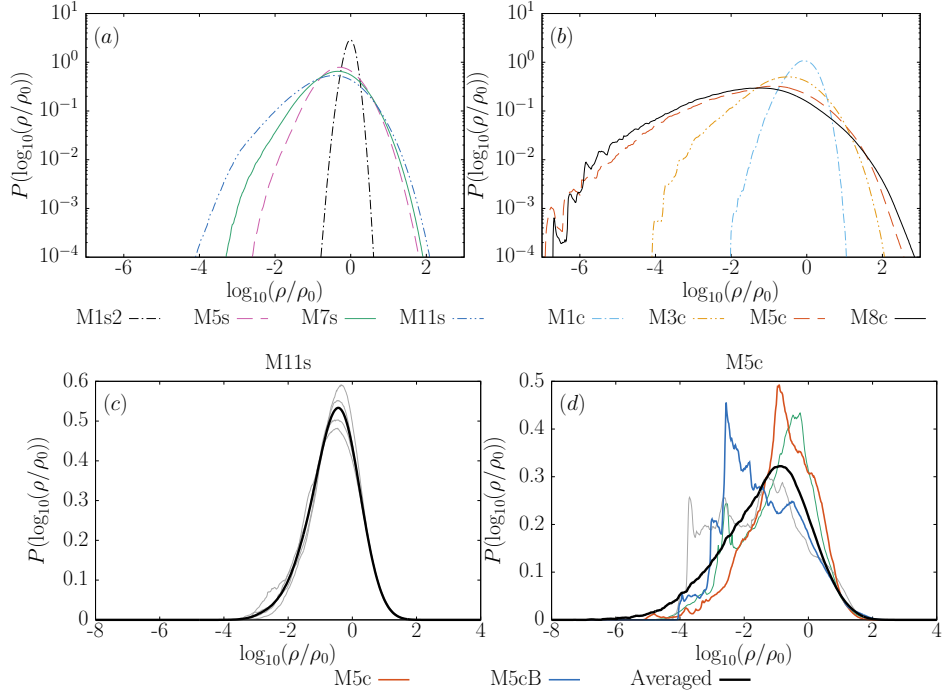
%
     \centering
{
	\vspace{\plotvspaceval em}
	\huge
 	\resizebox{1.\linewidth}{!}{\input{\grpathdenPDF_stats_jfm.tex}}
}
	\vspace{\plotvspacemid em}
{
	\huge
	\resizebox{1.\linewidth}{!}{\input{\grpathdenPDFvariations.tex}}
}

	\vspace{\plotvspacevalend em}

     \caption[Small caption]{\footnotesize{$(a\sm b)$ Mass density PDF of all transonic and supersonic runs, time-averaged over at least forty $512^3$ snapshots equally spaced in a time interval of about $4\ttt$. $(c\sm d)$ Snapshot-to-snapshot variations of the mass density PDFs for the M11s and M5c hydrodynamic steady-state. The thick line corresponds to the time average and the thiner lines to selected snapshots. Among them, the red and blue curves for the M5c steady-state correspond to different initial conditions for the magnetic helicity injection: the M5c run's initial condition corresponds to the red curve, whereas the M5cB run's initial condition corresponds to the blue one.}}

     \label{fig:denPDFvariations}
 \end{figure}

Figure \ref{fig:Hdenslices} shows mass density slices for the M1s2, M11s, M1c and M8c runs during the hydrodynamic steady-state. Since the mass density is governed by $\pat \rho=-\nabla \cdot (\rho \Fvel)$, a compressive forcing with a high $\nabla \cdot \Fvel$ component induces high mass density variations, visible through regions of very low density for the M8c run and pronounced shock fronts, already present at a RMS Mach number around unity (M1c run). Probability distribution functions (PDF) of the mass density statistics during the hydrodynamic steady-state are shown in figure \ref{fig:denPDFvariations} and display a considerably larger density spread for compressively-driven runs at relatively lower RMS Mach numbers as compared to the solenoidally-driven ones: for example, the spread of the M3c run is comparable to the M11s one. These curves are obtained by averaging over at least forty snapshots equally spaced in time over roughly $4\ttt$. A sampling rate higher than the turbulent turnover time is chosen because of the high variability of the density PDFs for the supersonic compressively-driven runs. For compressively-driven turbulence, the snapshot-to-snapshot variations of the density PDFs are significantly greater, as illustrated in figures \ref{fig:denPDFvariations}.$(c,d)$. For this reason, the influence of the particular snapshot taken as initial condition for magnetic helicity injection is assessed by taking two different instants in the M5c hydrodynamic steady-state as starting points. These runs are hereafter labelled respectively ``\MVcA'' and ``\MVcB'' and make use of the initial hydrodynamic state corresponding to the red (respectively blue) curve in figure \ref{fig:denPDFvariations}.$(d)$. For all the other runs, only one snapshot in the hydrodynamic steady-state is taken as initial condition for the magnetic helicity injection.

	For all the runs, the magnetic field-velocity field correlation,
\beq
\VBalign=\frac{\volavg{\Fvel \cdot \Fmag}}{\sqrt{\volavg{|\Fvel|^2} \volavg{|\Fmag|^2}}},
\eeq

grows in time but remains very low (under 0.035), so that the average alignment of magnetic and velocity field is small and the dynamics of the system are dominated by the direct cascades of kinetic and magnetic energies and the inverse transfer of magnetic helicity.

\label{place:mentionbottleneck}
Even though higher-order numerics allow to see comparatively many details as compared to lower-order schemes, the present simulations also exhibit well-known numerical inaccuracies such as, for example, a small-scale excess of fluctuations in the solution and the obvious consequences of limited spatial and temporal resolution (\cite{TEI20}, section 4.1.2). Section \ref{sec:confirmhighres} adresses this issue by testing the robustness of the obtained results.

\section{Notation}
\label{sec:notations}

The Fourier transform of an $\Lbox$-periodic function $\neutrfield$ is defined as:

\beq
\neutrfieldF_{\vk}=\frac{1}{N^3} \sum_{\vx} e^{-i\vk \cdot \vx} \neutrfield(\vx),
\eeq

with $\vx \in [0,\Lbox]^3$ the position vector, $\vk$ the wavenumber and $N$ the linear resolution of the cubic domain of size $\Lbox^3$ so that $N^3$ is the total number of cells. The power spectrum of $\neutrfield$ is defined as the shell-integrated spectrum:

\beq
	\label{eq:powerspec}
	\powerspec(\neutrfield)_\kspec=\frac{1}{2} \sum_{\kspec \leq |\vk|/\kone < \kspec+1} |\neutrfieldF_{\vk}|^2,
\eeq

with $\kone=2\upi/\Lbox$. On the other hand, a helicity spectrum is defined as the co-spectrum of a field $\neutrfield$ and its curl $\neutrfieldB=\nabla \times \neutrfield$:

\beq
	\label{eq:productspec}
	\productspec(\neutrfield,\neutrfieldB)_\kspec=\sum_{\kspec \leq |\vk|/\kone < \kspec+1} \neutrfieldF_{\vk} \cdot \neutrfieldBF_{\vk}^*,
\eeq

with the asterisk $(^*)$ designating the complex conjugate, so that $\powerspec(\neutrfield)=\frac{1}{2}\productspec(\neutrfield,\neutrfield)$.

The compressive part of a power spectrum considers only the contributions parallel to the wavevector $\vk$ and is defined as:

\beq	
	\label{eq:powerspeccomp}
	\powerspeccomp(\neutrfield)_\kspec=\frac{1}{2}\sum_{\kspec \leq |\vk|/\kone < \kspec+1} |\neutrfieldF_{\vk} \cdot \vk|^2/|\vk|^2.
\eeq

	The solenoidal part of a power spectrum is built through the difference $\powerspecsol(\neutrfield)=\powerspec(\neutrfield)-\powerspeccomp(\neutrfield)$.
	The Fourier spectra and co-spectra are furthermore normalised in order to refer more easily to physical quantities. The power spectrum of the velocity (the specific kinetic energy spectrum) and the kinetic helicity spectrum (co-spectrum of velocity and vorticity) are normalised by the isothermal sound speed squared $\cs^2$ so that $\powerspec(\Fvel)_\kspec=(1/2)\sum_{\kspec \leq |\vk|/\kone < \kspec+1} |\FvelF_{\vk} / \cs|^2$. Similarly, quantities involving the density, such as $\rhowIIv=\sqrt{\rho}\Fvel$ have an additional normalisation by an appropriate power of the mean density $\rhoz$. For example, $\powerspec(\rhowIIv)$ is normalised by $\rhoz \cs^2$. Since the mean density is $\rhoz=1$, this additional $\rhoz$ factor has no direct effect but is still mentioned since the magnetic field $\Fmag$ has the same dimension as $\rhowIIv$. For this reason, both the magnetic field power spectrum and the magnetic helicity spectrum are normalised by $\rhoz \cs^2$. These normalisation factors are implicitely assumed in the relations \eqref{eq:powerspec}-\eqref{eq:powerspeccomp} and not written explicitly in order to reduce the amount of notation.

	Furthermore, for the computation of the magnetic vector potential, the Coulomb gauge $\nabla \cdot \FmagA=0$ is chosen so that $\FmagAFk=i \vk \times \FmagF_{\vk}/k^2$ in Fourier space.

\section{Previous research}
\label{sec:prevresearch}
	An experimental setup similar to the one presented here has led to the observation of several spectral scaling laws for incompressible MHD turbulence \cite{MAL09,MMB12}. In these works, pseudo-spectral direct numerical simulations with $1024^3$ collocation points have been considered, with a magnetic helicity injection at the wavenumber shells $203\leq \kspec \leq 209$. Among other quantities, the kinetic helicity $\khelF=\productspec(\Fvel,\nabla \times \Fvel)$, magnetic helicity $\mhelF=\productspec(\FmagA,\Fmag)$, kinetic energy (which is the specific kinetic energy $\sekinF=\powerspec(\Fvel)$ in the incompressible case with a uniform $\rho=1$) and magnetic energy $\emagF=\powerspec(\Fmag)$ exhibit scaling laws $\sim \kspec^m$  with an exponent $m \approx -0.4, -3.3, -1.2$ and $-2.1$ respectively. Based on the Eddy Damped Quasi-Normal Markovian closure model (\cite{ORS70} extended to MHD in \cite{PFL76}), a dynamical equilibrium between shearing and twisting effects leads to the following relation \cite{MAL09,MMB12}:

\beq
\label{eq:Alfbalance}
\left(\frac{\sekinF}{\emagF}\right)^{\expoalf} \propto \frac{\khelF}{\jhelF},
\eeq

	with $\jhelF = \productspec(\Fmag,\nabla \times \Fmag)=(\frac{2\upi}{\Lbox}\kspec)^2 \mhelF$ the current helicity and $\expoalf$ a constant exponent, discussed below. This balance has also been observed (with $\expoalf=1$) in the direct transfer region of decaying MHD turbulence with a large scale helical magnetic field \cite{MIP09,GMP11} and has been interpreted as ``partial Alfv\'{e}nization of the flow'' \cite{GMP11}. It is henceforth referred to as ``Alfv\'{e}nic balance''.

	While the above-mentioned power-law exponents for the kinetic and magnetic energies and helicities are consistent with the Alfv\'{e}nic balance for $\expoalf=1$ \cite{MAL09}, a later work mentions that relation \eqref{eq:Alfbalance} is better verified with an exponent $\expoalf=2$ \cite{MMB12}. This could be due to the fact that the spectral domains where the scaling laws are followed are slightly different for the different quantities.

	The present work looks for scaling laws and aims at extending relation \eqref{eq:Alfbalance} to the compressible case.

\section{Results}
\label{sec:results}

\subsection{Structure formation}

\def\plotvspacevalend{-2}
	 
 \begin{figure}[h]
     \centering
     \huge
     \resizebox{1.\linewidth}{!}{\input{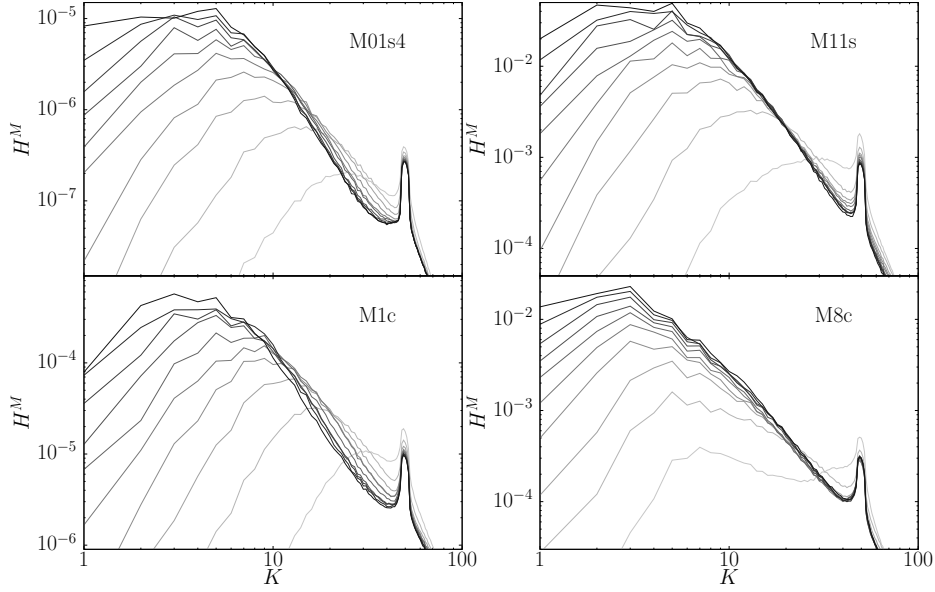}}	
     \caption[Small caption]{\footnotesize{Time evolution of the magnetic helicity spectra for the least and most compressible runs. The darker the curve, the later it is. The ten curves are equally-spaced in time and correspond to the time-intervals $t\in [0.335\ttt,3.35\ttt], [0.22\ttt,2.2\ttt], [0.24\ttt,2.4\ttt]$ and $[0.16\ttt,1.6\ttt]$ for the M01s4, M11s, M1c and M8c runs, respectively.}}
     \label{fig:invtsfMhl}
 \end{figure}

\def\plotvspacevalend{0}

The inverse transfer of magnetic helicity also takes place in supersonic flows, as shown in figure \ref{fig:invtsfMhl}, where the time evolution of the magnetic helicity spectra $\mhelF$ is shown for the least and most compressible cases considered. Figure \ref{fig:tevolIscHM} shows the time evolution of the magnetic helicity integral scale, defined as:

 \begin{figure}[h]
     \centering
     \huge
     \resizebox{1.\linewidth}{!}{\input{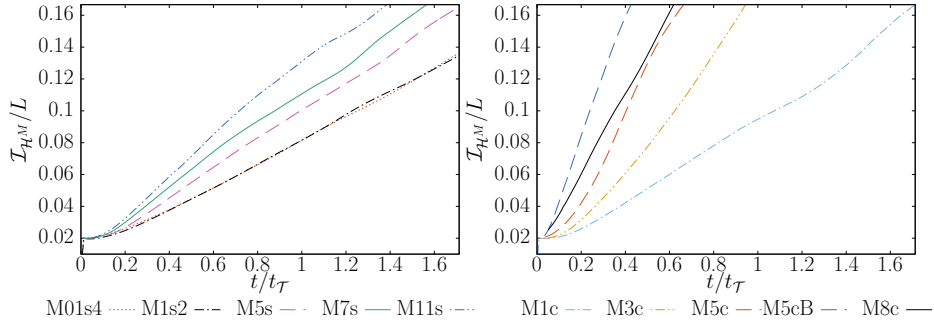}}	
     \caption[Small caption]{\footnotesize{Magnetic helicity integral scale time evolution.}}
     \label{fig:tevolIscHM}
 \end{figure}

\beq
	\label{eq:IscHm}
	\IscHm=\Lbox\frac{\int_{\kspec} \kspec^{-1} \mhelF_\kspec d\kspec}{\int_{\kspec} \mhelF_\kspec d\kspec}.
\eeq

Magnetic helicity is not a positive-definite quantity. Even though division by zero is in theory possible, since only magnetic helicity of one sign is injected, positive magnetic helicity dominates the system at all scales so that division by zero does not occur and considering this integral scale remains meaningful. The time is measured in units of the turbulent turnover time $\ttt=\Lbox/(2\cs\rmsM)$ of the hydrodynamic steady-state of the flow, so that $\rmsM$ designates the RMS Mach number before the injection of magnetic helicity. The inverse transfer occurs faster with increasing compressibility. While for the solenoidally-driven runs, more than $1.5\ttt$ are required to reach the integral scale $\IscHm=\frac{\Lbox}{6}$, less than about $0.6\ttt$ are needed for the compressively-driven runs with $\rmsM \gtrapprox 5$. For lower densities indeed, the Lorentz force has a greater impact on the velocity field's temporal evolution, since $\pat \Fvel = \ldots + \frac{1}{\rho} (\nabla \times \Fmag) \times \Fmag$, which backreacts on the magnetic field. At high compressibility, especially for compressively-driven runs, large regions of low density are present (figure \ref{fig:Hdenslices}), which are associated with higher Alfv\'{e}n velocities $\FvelA=\Fmag/\sqrt{\rho}$ and hence faster dynamical timescales of magnetic field line reconfiguration $\propto l_{rec}/\FvelA$, with $l_{rec}$ characterising the field-parallel extent of the magnetic reconnection region. Magnetic reconnection processes are required to achieve the topological changes of magnetic structures in the course of the inverse transport of magnetic helicity. Those changes are communicated along the affected field-lines by Alfv\'en-waves propagating at speed $\FvelA$. This explains why for the M5cB run, whose initial condition presents a peak of the density PDF at lower densities (figure \ref{fig:denPDFvariations}.$(d)$), the growth of the magnetic helicity integral scale is faster than for the M5c run. This is also manifest in the time evolution of the magnetic helicity spectra: for the M8c run, in contrast to the M01s4, M11s and M1c runs, the magnetic helicity spectrum goes first to significantly larger scales before steepening its spectral slope (figure \ref{fig:invtsfMhl}).

\subsection{Scaling relations: magnetic quantities}

	In order to compare the numerical models more easily with one another, the Fourier spectra of the different setups are considered at their respective instant in time when a magnetic helicity integral scale $\IscHm\approx\Lbox/6$ is reached. This scale is chosen in order to have a sufficiently large spectral distance between the magnetically forced and the largest scales, since a pollution of the scaling range by the electromotive forcing at small scales and through large-scale condensation of magnetic helicity at the largest scales at late times due to the limited box size are expected. The spectral power-law exponent is determined by a logarithmic linear least squares fit (LSF) to the spectrum in a relevant domain, which depends on the run and the considered quantity. This domain is shown through vertical dashed lines in the figures.

\newcommand{\tsMsMc}{\huge}
\newcommand{\vspaceEMsMc}{-1.}

\def\plotvspacevalend{\vspaceEMsMc}
	 
 \begin{figure}[h]
     \centering
     \tsMsMc
     \resizebox{1.\linewidth}{!}{\input{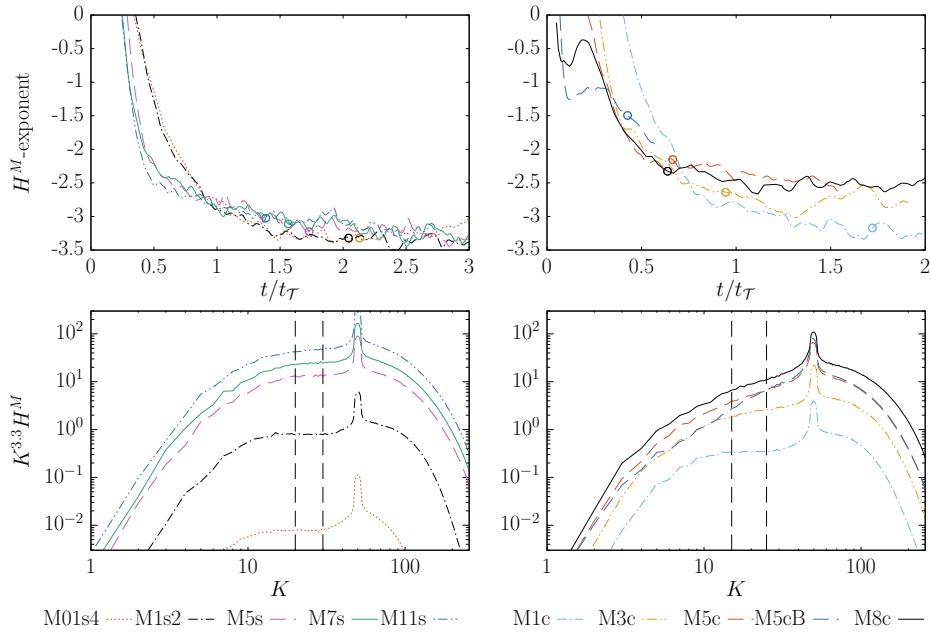}}	
     \caption[Small caption]{\footnotesize{Top: Time evolution of the magnetic helicity power-law exponents. Bottom: magnetic helicity spectra at an instant when $\IscHm\approx\frac{\Lbox}{6}$ (shown through circles in the top plots), compensated by $\kspec^{3.3}$, since $\mhelF \sim \kspec^{-3.3}$ in the incompressible case. The exponents measured in the top plots are gained through a LSF in the region delimited with vertical dashed lines in the bottom plots (here $20 \leq \kspec \leq 30$ and $15 \leq \kspec \leq 25$ for the solenoidally- and compressively-driven runs respectively).}}
     \label{fig:HmSpec}
 \end{figure}

\def\plotvspacevalend{0}

As expected from the presence of $\nabla \cdot \Fvel$ terms in the nonlinear evolution equations of magnetic helicity and magnetic energy, the respective spectra display systematic changes when augmenting the Mach number of the flow. Those are more pronounced for compressively-driven turbulence than for solenoidally-driven flows. The low Mach number solenoidally-driven M01s4 and M1s2 runs exhibit an approximate magnetic helicity power-law scaling $\mhelF \sim \kspec^m$ with $m \approx -3.3$, gained from a LSF in the region $20 \leq \kspec \leq 30$ (figure \ref{fig:HmSpec}). This exponent is consistent with the observations in the incompressible case \cite{MAL09,MMB12}. For higher RMS Mach numbers, increasing deviations are observed, namely $m\approx -3.2, -3.1$ and $-3.0$ for the M5s, M7s and M11s runs respectively. Although significant, these deviations are clearly smaller than those obtained in the compressively-forced runs. There, the spectra are significantly flatter at lower Mach numbers, see figure \ref{fig:HmSpec}. For the M1c, M3c, M5c and M8c runs, the exponents computed through a LSF in the region $15 \leq \kspec \leq 25$ are $m \approx -3.2, -2.6, -2.2$ and $-2.3$ respectively. For supersonic compressively-driven flows, the scaling exponent is very sensitive to the initial conditions: for example, the M5cB run, for which the magnetic helicity injection starts at a different instant during the same hydrodynamic statistically steady-state as the M5c run, presents a significantly flatter spectrum with $m \approx -1.5$. This is linked with a peak of the initial condition's mass density PDF at lower densities (figure \ref{fig:denPDFvariations}.$(d)$), so that faster timescales are expected for the M5cB run. This illustrates that, if one would aim at determining a typical magnetic helicity scaling exponent dependency as a function of $\rmsM$ using a compressive forcing, one would need to average the measures over many realisations, which is not in the scope of the present work. Nevertheless, a clear tendency towards flatter exponents with increasing compressibility is observable, which is alleviated by an appropriate choice of variables, even with very different initial conditions, as shown below.

	The same tendency is observed for the magnetic energy spectra: the scaling exponents are close to the $-2.1$ observed in the incompressible case \cite{MAL09,MMB12} for the low Mach number M01s4, M1s2 and M1c runs, flatter with increasing compressibility for the solenoidally-driven runs and significantly flatter for the compressively-driven runs already at lower initial RMS Mach numbers (curves not shown here).%

	In the incompressible case, the Alfv\'{e}n velocity $\FvelA=\Fmag/\sqrt{\rho}$ (with $\rho$ a constant) and the magnetic field are in essence the same quantity. However, in the compressible case, low density regions are associated with higher Alfv\'{e}n velocities, so that the dynamical timescales are shorter there. For this reason, it is tempting to consider the power spectra of the Alfv\'{e}n velocity $\ealfF=\powerspec(\FvelA)$ as well as the co-spectrum of the Alfv\'{e}n velocity and its curl $\ahelF=\productspec(\FvelA,\nabla \times \FvelA)$, which is the helicity of the Alfv\'{e}n velocity, a quantity hereafter called ``Alfv\'{e}nic helicity''. In the incompressible case, the Alfv\'{e}nic helicity is $\Fmag \cdot \FmagJ/\rho$ and corresponds thus to the current helicity $\jhelF = (\frac{2\upi}{\Lbox}\kspec)^2 \mhelF$, which exhibits an approximate scaling law $\jhelF \sim \kspec^{-1.3}$ \cite{MAL09,MMB12}.

\newcommand{\sameasfig}[2]{Same as figure \ref{fig:HmSpec}, but #1 which exhibits a #2 scaling in the incompressible case.}

\def\plotvspacevalend{\vspaceEMsMc}
	 
 \begin{figure}[h]
     \centering
     \tsMsMc
     \resizebox{1.\linewidth}{!}{\input{\grpathspec_lsf_Ha_Ms_Mc_jfm.tex}}	
     \caption[Small caption]{\footnotesize{\sameasfig{for the Alfv\'{e}nic helicity $\ahelF$}{$\kspec^{-1.3}$}}}
     \label{fig:HaSpec}
 \end{figure}

\def\plotvspacevalend{0}

\begin{figure}[h]
     \centering
     \huge
     \resizebox{1.\linewidth}{!}{\input{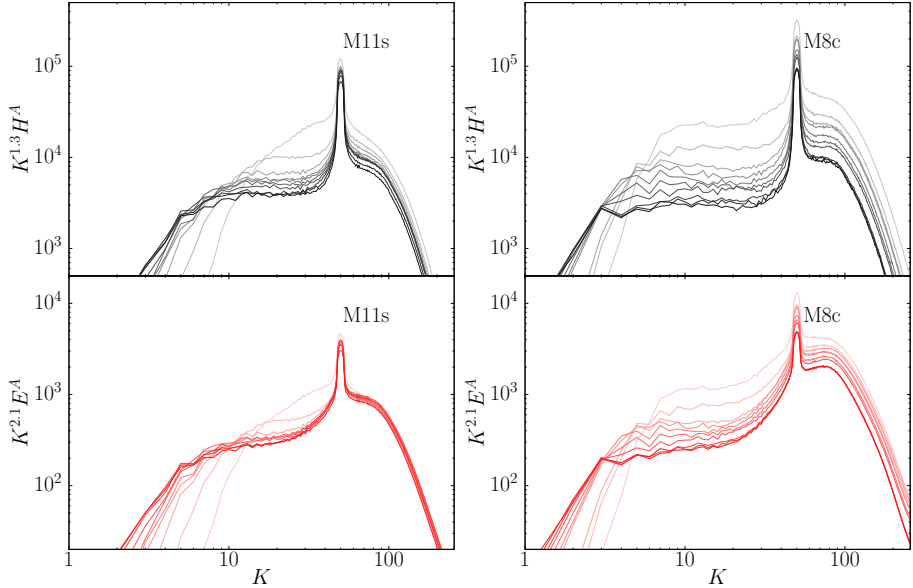}}	
     \caption[Small caption]{\footnotesize{Time evolution of $\ahelF$ and $\ealfF$ for the M11s and M8c runs, compensated by $\kspec^m$ with $-m$ the scaling exponent in the incompressible case. The spectra are snapshots taken at different instants (the darker, the later), from $t=t_f/10$ to $t_f=2.2\ttt$ and $t_f=1.6\ttt$ for the M11s and M8c runs respectively.}}
     \label{fig:HEatime}
 \end{figure}

	The Alfv\'{e}nic helicity and Alfv\'{e}n velocity power spectra display only a weak dependence on the compressible character of the forcing and the flow. While the magnetic energy and helicity spectra become flatter with increasing compressibility, for the Alfv\'{e}nic helicity the incompressible scaling exponent $-1.3$ is approximately observed for all simulations with $\rmsM \gtrsim 3$ as well, see figure \ref{fig:HaSpec}). This is in particular the case for the M5c and M5cB runs, which present very different $\mhelF$ scaling exponents. The scaling exponents of $\ealfF$, which approach asymptotically $-1.8$ for the most compressible runs (see figure \ref{fig:HEatime}), are also closer to the $-2.1$ scaling for the magnetic energy in the incompressible case \cite{MAL09,MMB12} and present a weaker dependence on the flow's compressibility as compared to $\emagF$.

\def\plotvspacevalend{-2}
	 
\def\plotvspacevalend{0}

\def\plotvspacemid{0}
	 
 \begin{figure}
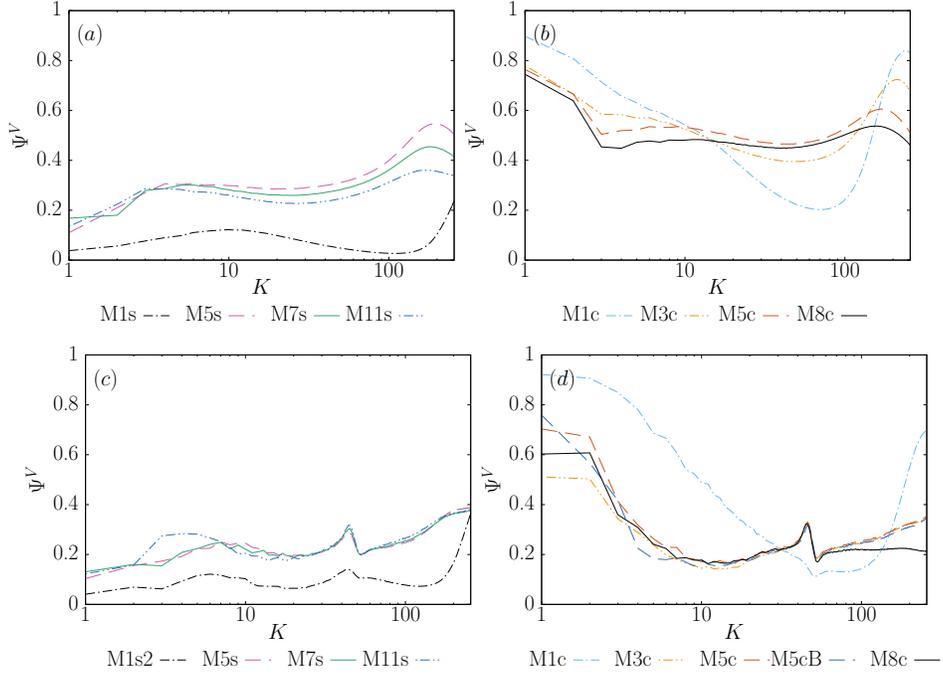
%
     \centering
{
	\vspace{\plotvspaceval em}
	\huge
 	\resizebox{1.\linewidth}{!}{\input{\grpathcompratioV_hydro_jfm.tex}}
}
	\vspace{\plotvspacemid em}
{
	\huge
	\resizebox{1.\linewidth}{!}{\input{\grpathcompratio_Ms_Mc_onlyV_jfm.tex}}
}

	\vspace{\plotvspacevalend em}

     \caption[Small caption]{\footnotesize{$(a\sm b)$ Time-averaged velocity compressive ratio during the hydrodynamic steady-state, exhibiting values close to 1/3 for the solenoidally-driven supersonic runs and 1/2 for the supersonic compressively-driven ones in a reasonable wavenumber range, consistently with \cite{FRD10}. $(c\sm d)$ Snapshots of the compressive ratio at an instant when $\IscHm\approx\Lbox/6$: all supersonic runs exhibit a compressive ratio close to 0.2, independently of the mechanical forcing type. The M1c run continues to exhibit a large compressive ratio.}}

     \label{fig:compratioV}
 \end{figure}

\def\plotvspacemid{0}

	This universality of the exponents over a wide range of compressibility suggests a systematic scale-dependent correlation between the density and the magnetic fields, which is the object of further investigations. The M1c run may present different dynamics since its velocity compressive ratio $\powerspeccomp(\Fvel)/\powerspec(\Fvel)$ remains large in the inverse transfer region (see figure \ref{fig:compratioV}). This is reflected by the lack of low density regions in this simulation.

\subsection{Scaling relations: kinetic quantities}

\def\plotvspacevalend{\vspaceEMsMc}
	 
 \begin{figure}[h]
     \centering
     \tsMsMc
     \resizebox{1.\linewidth}{!}{\input{\grpathspec_lsf_EkVs_Ms_Mc_jfm.tex}}	
     \caption[Small caption]{\footnotesize{\sameasfig{for the solenoidal part of the specific kinetic energy spectrum $\sekinFsol$}{$\kspec^{-1.2}$}}}
     \label{fig:EkVsSpec}
 \end{figure}

\def\plotvspacevalend{0}

	In the incompressible case, the kinetic energy and helicity exhibit approximate scaling laws with exponents close to $-1.2$ and $-0.4$ respectively \cite{MAL09,MMB12}. In the compressible case, several alternative definitions for the incompressible kinetic energy have been considered in the literature, e.g. $\sekinF=\powerspec(\Fvel)$, i.e. the specific kinetic energy spectrum, $\ekinF=\powerspec(\rhowIIv)$, with $\rhowIIv=\sqrt{\rho}\Fvel$, which is dimensionally a kinetic energy spectrum, $\EkRF=\powerspec(\rhowIIIv)$, with $\rhowIIIv=\rho^{1/3}\Fvel$, a quantity for which the Kolmogorov spectrum with exponent $-\frac{5}{3}$ is recovered in compressible hydrodynamic turbulence with a low $\nabla \cdot \Fvel$ component \cite{FLE96,KNP07,FRD10,FED13}, $\frac{1}{2}\productspec(\rho\Fvel,\Fvel)$, the co-spectrum of velocity and momentum (or more generally, any co-spectrum $\frac{1}{2}\productspec(\rho^{\alpha}\Fvel,\rho^{1-\alpha}\Fvel)$ with $\alpha \in [0,1]$ \cite{ALU13}), or $\sekinFsol=\powerspecsol(\Fvel)$, the solenoidal part of the specific kinetic energy spectrum.

\def\plotvspaceval{-12}
\def\plotvspacevalend{0}
\def\plotvspacemid{0}
	 
 \begin{figure}%
     \centering
{
	\vspace{\plotvspaceval em}
	\tsMsMc
 	\resizebox{1.\linewidth}{!}{\input{\grpathspec_lsf_Hk_Ms_Mc_jfm_edit.tex}}
}
	\vspace{\plotvspacemid em}
{
	\Huge
	\resizebox{0.6\linewidth}{!}{\input{\grpathMm_HEv.tex}}
}

	\vspace{\plotvspacevalend em}

     \caption[Small caption]{\footnotesize{Top: Kinetic helicity spectra $\khelF$ at an instant when $\IscHm=\frac{\Lbox}{6}$ compensated by $\kspec^{0.4}$. Bottom: Time evolution of $\khelF$, compensated by $\kspec^{0.4}$, for a low Mach number run without large-scale mechanical driving, with $\EMinj=7\times 10^{-4}$ (for which $\rmsM \approx 0.2$). The curves are equally spaced in time from an instant when $\IscHm\approx\Lbox/19$ till $\IscHm \approx \Lbox/3$. The darker the curve, the larger $\IscHm$.}}

     \label{fig:HvSpec}
 \end{figure}

\def\plotvspaceval{0}
\def\plotvspacemid{0}
\def\plotvspacevalend{0}
\label{sec:manyEkdef}
From the above-mentioned quantities, the specific kinetic energy spectrum $\sekinF=\powerspec(\Fvel)$ and its solenoidal part $\sekinFsol$ (shown in figure \ref{fig:EkVsSpec}) present the most universal behaviour, with an exponent close to $-1.2$ for all the runs. The deviations from the $-1.2$ exponent become weaker at later times for the M01s4 and M1s2 runs and the stronger deviations for the M1c run are due to the high velocity spectrum compressive ratio. The other above-mentioned spectra are displayed in appendix \ref{app:Ekspec}.

	As for the kinetic helicity, the $-0.4$ incompressible exponent appears to be the asymptotic behaviour for all levels of higher compressibility (figure \ref{fig:HvSpec}). The spectral transient region influenced by the small-scale magnetic driving is wider at lower Mach numbers. This is probably linked with the presence of a large-scale mechanical driving in the present experiments, as compared to those of \cite{MMB12}. In hydrodynamic turbulence, the energy cascade tends to an equipartition of energy in the positive and negative helical parts of the velocity field at small scales, so that the kinetic helicity tends to zero \cite{CCE03}. On the other hand, the injection of magnetic helicity at small scales leads to the generation of kinetic helicity. It is plausible that for runs with pronounced compressible effects and low density regions, the dynamical timescale of the kinetic helicity production is faster than that of the direct energy cascade leading to vanishing kinetic helicity, so that the $-0.4$ exponent observed in the absence of large-scale mechanical driving is recovered, but only for the supersonic flows. As shown in figure \ref{fig:HvSpec}, bottom, a low Mach number run performed in the absence of a large-scale mechanical driving gives a kinetic helicity scaling compatible with $-0.4$.

\subsection{Alfv\'{e}nic balance}
\label{sec:Alfvenicbalance}
The observations presented so far suggest that the Alfv\'{e}nic balance found in the incompressible case (relation \eqref{eq:Alfbalance}) can be extended to the compressible case when considering, instead of the current helicity $\jhelF$ and the magnetic energy $\emagF$, the Alfv\'{e}nic helicity $\ahelF$ and the power spectrum of the Alfv\'{e}n velocity $\ealfF$ respectively, and by considering the specific kinetic energy $\sekinF$ or its solenoidal part. Since the energy associated with $\ahelF$ and the kinetic helicity $\khelF$ is the solenoidal part of $\ealfF$ and of $\sekinF$, respectively, the following extension of the Alfv\'{e}nic balance in the compressible case is proposed:

\beq
\label{eq:Alfbalancecomp}
\left(\frac{\sekinFsol}{\ealfFsol}\right)^{\expoalf} \propto \frac{\khelF}{\ahelF}.
\eeq

This choice is a straightforward extension of the balance in the incompressible case, where only solenoidal modes exist. To test this relation, the ratio

\begin{figure}[h]
     \centering
     \huge
     \resizebox{1.\linewidth}{!}{\input{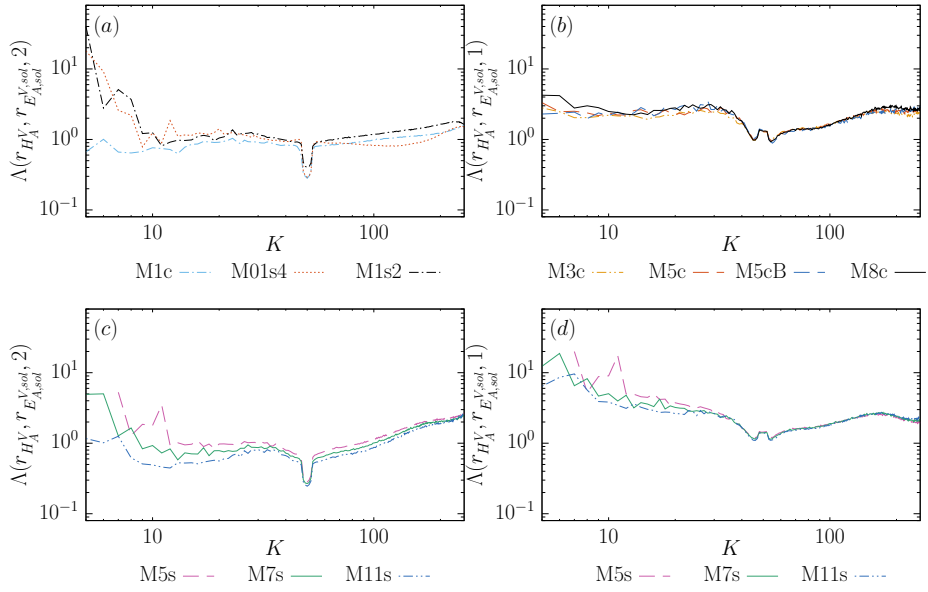}}	
     \caption[Small caption]{\footnotesize{Alfv\'{e}nic balance for different runs and different exponents $\expoalf$: $(a)$ For the subsonic and transonic runs, $\gamma=2$, compatible with the results in the incompressible case. $(b)$ For the supersonic compressively-driven runs, which verify the Alfv\'{e}nic balance for $\gamma=1$. $(c\sm d)$ For the supersonic solenoidally-driven runs, which present an intermediate behaviour with respect to the $\gamma$ exponent.}}
     \label{fig:Alfvbalance}
 \end{figure}

\newcommand{\LambH}{r_H}
\newcommand{\LambHVA}{r_{H^V_A}}
\newcommand{\LambHVJ}{r_{H^V_J}}
\newcommand{\LambE}{r_E}
\newcommand{\LambEVAsol}{r_{E^{V,sol}_{A,sol}}}
\newcommand{\LambEKA}{r_{E^{K}_{A}}}
\newcommand{\LambEKM}{r_{E^{K}_{M}}}

\beq
	\label{eq:Alfvariants}
	\Lambda(\LambH,\LambE,\gamma)=\frac{\LambE^\gamma}{\LambH},
\eeq

with $\LambH=\LambHVA=\khelF/\ahelF$ and $\LambE=\LambEVAsol=\sekinFsol/\ealfFsol$ is plotted in figure \ref{fig:Alfvbalance} for the different runs and different $\expoalf$. This relation is well followed for the least compressible M01s4, M1s2 and M1c runs with $\expoalf=2$, with $\Lambda \approx 1.1 \pm 0.3$ on the domain $20 \leq \kspec \leq 44$, consistently with \cite{MMB12}. For the most compressible cases, the curves for the M3c, M5c, \MVcB\ and M8c runs show a similar behaviour for $\expoalf=1$, with $\Lambda \approx 2.7 \pm 0.5$ on the $13 \leq \kspec \leq 33$ domain. Regarding the supersonic solenoidally-driven runs, they present an intermediate behaviour: for $\expoalf=2$, the M5s curve would be included in the ``M01s4-M1s2-M1c bundle'' on the domain $12 \leq K \leq 40$, whereas for $\expoalf=1$, the M11s curve would be included in the ``M3c-M5c-M8c bundle'' on the domain $16 \leq \kspec \leq 31$.

Replacing the solenoidal parts $\sekinFsol$ and $\ealfFsol$ by the total energies $\sekinF$ and $\ealfF$ in $\Lambda$ also gives a good horizontal for the high Mach number compressively-driven runs. The ``M3c-M5c-M8c bundle'' verifies then $\Lambda \approx 3 \pm 0.5$ on the same $13 \leq \kspec \leq 33$ domain. With this choice however, the M1c curve does not stay in the ``M01s4-M1s2'' bundle, which is why the choice of relation \eqref{eq:Alfbalancecomp} is preferred.

Although other variants of relation \eqref{eq:Alfbalancecomp} are partially consistent with the data as well, this choice shows overall the best agreement with the simulations, as well as the least spread between the curves. This also holds at higher numerical resolution, which is not the case for some other possibilities, as shown in section \ref{sec:confirmhighres}.

\newcommand{\alfJFMVAsolI}{\Lambda(\LambHVA,\LambEVAsol,1)}
\newcommand{\alfJFMVAsolII}{\Lambda(\LambHVA,\LambEVAsol,2)}
\newcommand{\alfJFMKAI}{\Lambda(\LambHVA,\LambEKA,1)}
\newcommand{\alfJFMKAII}{\Lambda(\LambHVA,\LambEKA,2)}
\newcommand{\alfJFMKMI}{\Lambda(\LambHVJ,\LambEKM,1)}
	 
\section{Confirmation at higher resolution}
\label{sec:confirmhighres}
	In addition to the already mentioned inaccuracies of the present numerical experiments, in the framework of the fallback approach, not all cells are reconstructed at higher order. For the most compressible M11s and M8c runs, a little more than 50\% of the reconstructions occur indeed at third order or lower. It is hence important to check that the results presented here are reasonably well converged, both with respect to the resolution and the numerical scheme's order. Therefore, some additional runs are considered: the M1s2LR and M1s2HR runs, which are lower resolution ($256^3$) and higher resolution ($1024^3$) runs respectively and the M1s2LO run which is done at resolution $512^3$ but using a second-order scheme. Similarly, additional runs for the M8c case are considered, labelled analogously ``M8cLR'', ``M8cHR'' and ``M8cLO''. For this supersonic compressively-driven run, the timestep $\Dt$ becomes very small at higher resolutions, due to very high Alfv\'{e}n speeds in low density regions, so that only an early instant in time is considered, when $\IscHm \approx 0.045 \Lbox$, but where scaling behaviour is already observed. %

\begin{figure}[h]
     \centering
     \Huge
     \resizebox{1.\linewidth}{!}{\input{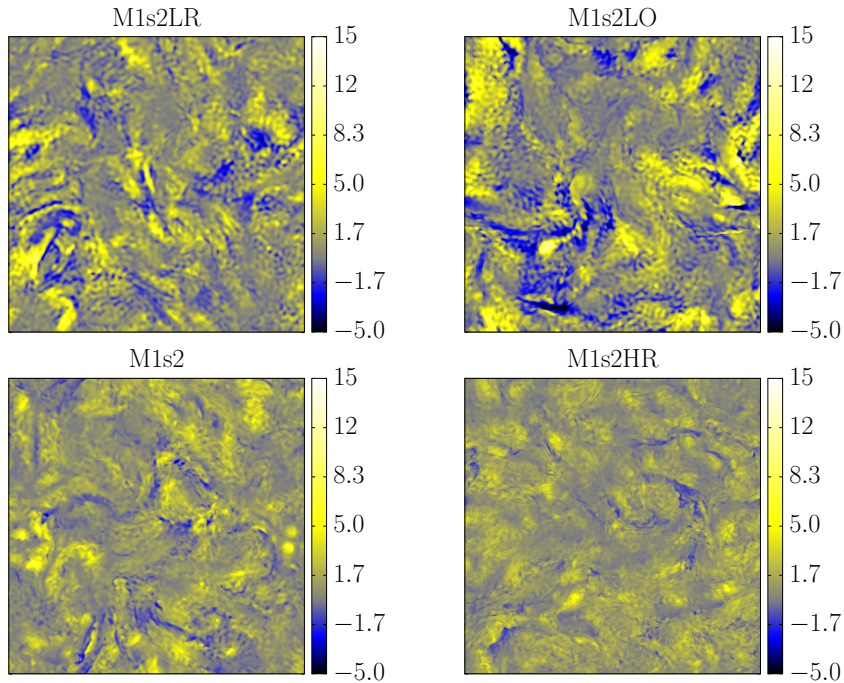}}	
     \caption[Small caption]{\footnotesize{Magnetic helicity density slices, normalised by the mean magnetic helicity in the system, at an instant when $\IscHm\approx \frac{1}{6}\Lbox$.}}
     \label{fig:lowerorder_hmslices}
 \end{figure}

	Finer structures are visible for the M1s2 family in the magnetic helicity density slices with increasing resolution and/or using a higher-order scheme (figure \ref{fig:lowerorder_hmslices}), as a result of accuracy gain and significant reduction of numerical dissipation. The level of detail of the $512^3$ lower-order run is very similar to the one of the $256^3$ higher-order run, showing the benefits of using higher-order numerics: even though they are computationally more expensive at a given resolution, they allow a good result's accuracy at comparatively lower resolution, resulting in a net performance gain.

\begin{figure}[h]
     \centering
     \huge
     \resizebox{1.\linewidth}{!}{\input{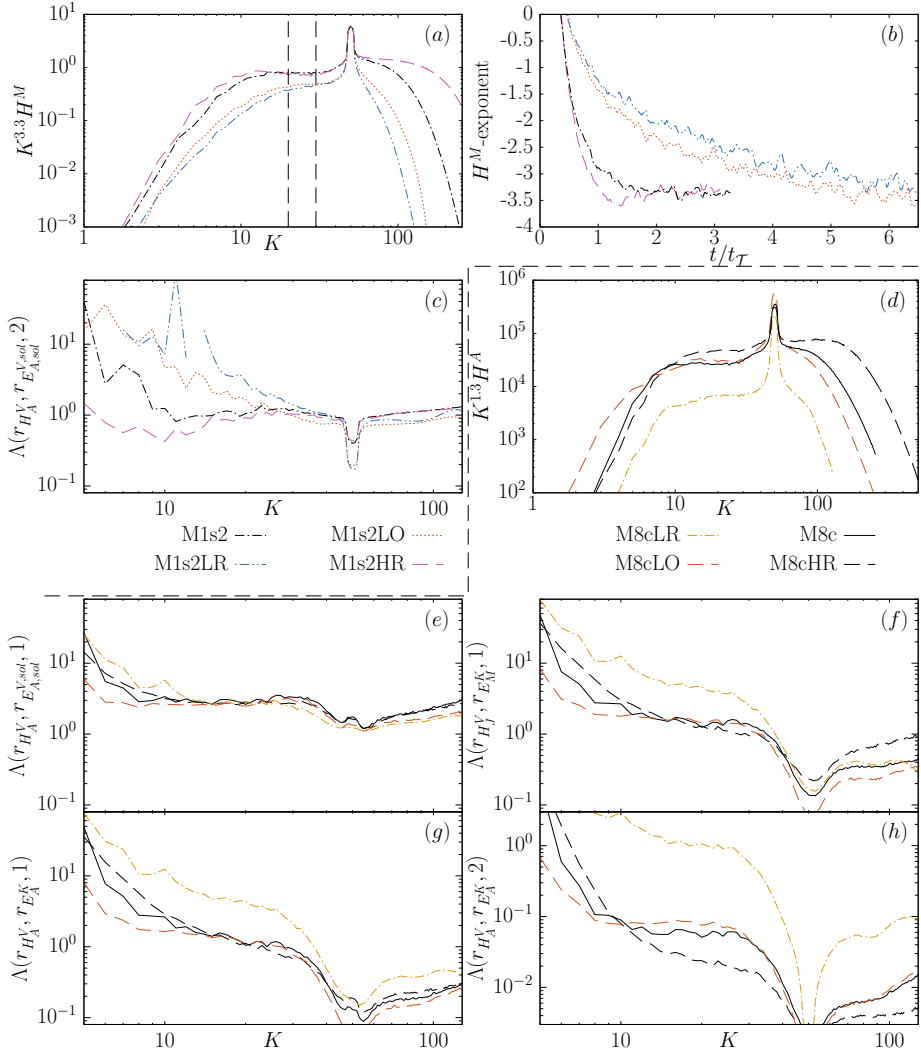}}	
     \caption[Small caption]{\footnotesize{Confirmation of results through a resolution/numerical order study. The $(a\sm c)$ subfigures above the dashed line are for the M1s2 family, the $(d\sm h)$ subfigures under the line for the M8c family. $(a)$ Compensated magnetic helicity spectra for the M1s2 family, the dashed vertical lines delimiting $20 \leq \kspec \leq 30$ show the domain where the LSF occurs in subfigure $(b)$, which shows the time evolution of the magnetic helicity scaling exponent. $(c)$ Verification of the Alfv\'{e}nic balance. $(d)$ Compensated Alfv\'{e}nic helicity spectra for the M8c family. $(e \sm h)$ Different Alfv\'{e}nic balance variants (see relation \eqref{eq:Alfvariants}) for the M8c family, with $\LambHVJ=\khelF/\jhelF$ and similarly $\LambEKA=\ekinF/\ealfF, \LambEKM=\ekinF/\emagF$.}}
     \label{fig:convstudy}
 \end{figure}

For the M1s2 family, a scaling exponent consistent with $\mhelF \sim \kspec^{-3.3}$ is observed as well for the higher resolution run (figure \ref{fig:convstudy}.$(a)$). The lower resolution and lower order runs also seem to converge to this exponent, but more slowly (figure \ref{fig:convstudy}.$(b)$). The Alfv\'{e}nic balance (relation \eqref{eq:Alfbalancecomp}) is also well verified with $\expoalf=2$ (figure \ref{fig:convstudy}.$(c)$). Similarly, for the M8c family, the scaling exponents for the Alfv\'{e}nic helicity are consistent with $\ahelF \sim \kspec^{-1.3}$ for all the runs and the Alfv\'{e}nic balance \eqref{eq:Alfbalancecomp} looks well converged for $\expoalf=1$ (figures \ref{fig:convstudy}.$(d,e)$). Other variants, shown in figures \ref{fig:convstudy}.$(f \sm h)$ with different choices for $(\LambH,\LambE,\gamma)$ in relation \eqref{eq:Alfvariants}, which also present a horizontal line at resolution $512^3$, present deviations from the horizontal at the higher $1024^3$ resolution as well as a greater resolution dependence. This hints to a better validity of the variant expressed in relation \eqref{eq:Alfbalancecomp}.

	These observations suggest that even though studies at higher resolutions would be beneficial to capture all the turbulent fine structures, the observed scaling laws are quite robust.

\section{Conclusion}
\label{sec:conclusion}

Similarly to the incompressible case, spectral scaling laws are observed for the magnetic helicity inverse transfer in supersonic compressible isothermal MHD turbulence. Magnetic helicity is injected in large-scale mechanically-driven hydrodynamic turbulent steady-states, with RMS Mach numbers ranging from about 0.1 to 11 for a purely solenoidal forcing and from about 1 to 8 for a purely compressive driving. The higher-order numerics allow results of convincing accuracy already at resolution $512^3$: in spite of well-known numerical inaccuracies, the main results seem well converged when comparing to $256^3$ and $1024^3$ runs.

The scaling exponents obtained are generally in good agreement with those found in the incompressible case for subsonic and transonic solenoidally-driven flows. In particular, the magnetic helicity spectrum goes as $\kspec^{m}$ with $m \approx -3.3$. The spectra become increasingly flatter with higher compressibility. The deviations are however relatively small for the solenoidally-driven runs even at RMS Mach numbers of the order of ten as compared to the purely compressive driving, where the spectra are significantly flatter already at a RMS Mach number around three in the initial hydrodynamic steady-state.

In the interstellar medium, both cases of a purely solenoidal or a purely compressive forcing are very unlikely to happen \cite{FRD10}. This shows nevertheless that effects of compressibility are expected to have a significant impact on the magnetic helicity dynamics in astrophysical systems of interest already at relatively low RMS Mach numbers, especially in situations where the turbulence drivers are rather compressive.

An appropriate change of variable, namely considering the Alfv\'{e}n velocity $\FvelA=\Fmag/\sqrt{\rho}$ instead of the magnetic field, alleviates the differences between the compressible and the subsonic/transonic solenoidally-driven runs. The co-spectrum of the Alfv\'{e}n velocity and its curl, named here ``Alfv\'{e}nic helicity'' presents for all the runs but the M1c one a similar scaling, consistent with the incompressible case one, going as $\kspec^{m}$ with $m \approx -1.3$. This points out to a certain universality in the inverse transfer dynamics, valid over a wide range of compressibility.

Lastly, the ``Alfv\'{e}nic balance'' (a quasi-equipartition in terms of magnetic and kinetic energies and helicities) observed in incompressible MHD has been extended using an appropriate change of variables. Further investigations of the balance relation are under way.

The authors acknowledge the North-German Supercomputing Alliance (HLRN) for providing HPC resources that have contributed to the research results reported in this paper. Computing resources from the Max Planck Computing and Data Facility (MPCDF) are also acknowledged. JMT gratefully acknowledges support by the Berlin International Graduate School in Model and Simulation based Research (BIMoS).

Declaration of Interests. The authors report no conflict of interest.

\appendix

\section{Spectra linked with kinetic energy}
\label{app:Ekspec}

As mentioned in section \ref{sec:manyEkdef}, the (specific) kinetic energy spectrum in the incompressible case where the mass density $\rho$ is constant can correspond to several quantities in the compressible case. Figure \ref{fig:EkVRMSpec} shows some possibilities not displayed in the main text, at an instant when $\IscHm \approx \Lbox/6$. The spectra are compensated by $\kspec^{1.2}$ since the scaling observed in the incompressible case is $\sekinF=\powerspec(\Fvel) \sim \kspec^{m}$ with $m\approx -1.2$ \cite{MMB12}. This $\sekinF$ scaling is well verified in the supersonic flows studied here as well. Deviations are observed when considering $\powerspec(\rho^{1/3}\Fvel)$, $\powerspec(\rho^{1/2}\Fvel)$ or the co-spectrum $\frac{1}{2}\productspec(\rho \Fvel,\Fvel)$.

\newcommand{\EkVvarnameJFM}{\kspec^{1.2}\powerspec(\Fvel)}
\newcommand{\EkvarnameJFM}{\kspec^{1.2}\powerspec(\rho^{1/2}\Fvel)}
\newcommand{\EkRvarnameJFM}{\kspec^{1.2}\powerspec(\rho^{1/3}\Fvel)}
\newcommand{\EkMvarnameJFM}{\frac{1}{2}\kspec^{1.2}\productspec(\rho\Fvel,\Fvel)}
	 
 \begin{figure}[h]
     \centering
     \huge
     \resizebox{1.\linewidth}{!}{\input{\grpathspec_EkVRM_jfm.tex}}	
     \caption[Small caption]{\footnotesize{Different variants corresponding to the (specific) kinetic energy in the incompressible case (where $\powerspec(\Fvel) \sim \kspec^{-1.2}$), taken at an instant when $\IscHm \approx \Lbox/6$.}}
     \label{fig:EkVRMSpec}
 \end{figure}

\bibliography{biblio}
\bibliographystyle{plain}

\end{document}

%% file: includes/newcommands.tex
\input{nc_num}

\input{nc_fourier}

\input{nc_tilde}
\input{nc_arrows}
\input{nc_domain}
\input{nc_eq}

\input{nc_vec}

\input{nc_bar}
\input{nc_misc}

\input{nc_physics}

\input{nc_experiments}

%% file: includes/nc_num.tex
\newcommand{\Dt}{\Delta t}

%% file: includes/nc_fourier.tex
\newcommand{\vhf}{\vec{\hat {f}}}
\newcommand{\vhg}{\vec{\hat {g}}}


\newcommand{\vhhs}[2]{\vec{{{\hat{h}}}^{#1}_{#2}}}

%% file: includes/nc_eq.tex
\newcommand{\beqa}{\begin{eqnarray}}
\newcommand{\eeqa}{\end{eqnarray}}
\newcommand{\beq}{\begin{equation}}
\newcommand{\eeq}{\end{equation}}

\newcommand{\beqal}{\begin{equation}\begin{aligned}}
\newcommand{\eeqal}{\end{aligned}\end{equation}}

\newcommand{\pat}{\partial_t}

%% file: includes/nc_vec.tex
\newcommand{\vE}{\vec{E}}

\newcommand{\vf}{\vec{{f}}}

\newcommand{\vg}{\vec{{g}}}

\newcommand{\vk}{\vec{k}}

\newcommand{\vN}{\vec{N}}

\newcommand{\vW}{\vec{W}}

\newcommand{\vx}{\vec{x}}

%% file: includes/nc_misc.tex
\newcommand{\bit}{\begin{itemize}}
\newcommand{\eit}{\end{itemize}}

%% file: includes/nc_physics.tex

\newcommand{\neutrAmp}{A}
\newcommand{\neutrfield}{\vf}
\newcommand{\neutrfieldB}{\vg}
\newcommand{\neutrfieldF}{\vhf}
\newcommand{\neutrfieldBF}{\vhg}

\newcommand{\solexpo}{sol}

\newcommand{\compexpo}{comp}

\newcommand{\neutrenergysymb}{{\cal{E}}}

\newcommand{\rmsM}{{\cal{M}}} 
\newcommand{\kspec}{K} 



\newcommand{\hel}{{\cal{H}}}
\newcommand{\mhel}{{\hel^M}} 

\newcommand{\ahelF}{H^A} 
\newcommand{\jhelF}{H^J} 
\newcommand{\mhelF}{H^M} 
\newcommand{\khelF}{H^V} 

\newcommand{\ealfF}{E^A} 
\newcommand{\ealfFsol}{E^{A,\solexpo}} 
\newcommand{\emagF}{E^M} 


\newcommand{\ekinF}{E^K} 



\newcommand{\sekinF}{E^V} 

\newcommand{\sekinFsol}{E^{V,\solexpo}} 


\newcommand{\famp}{F_0}

\newcommand{\fMAGk}{\vec{F^M_{\vk}}}

\newcommand{\fOUF}{\vec{F^{OU}}}
\newcommand{\fOUFk}{\vec{ F^{OU}_{\vk}}}
\newcommand{\fat}{t_{auto}} 
\newcommand{\ttt}{t_{\cal{T}}} 
\newcommand{\Lbox}{L} 

\newcommand{\Einj}{{\epsilon^{K}_{inj}}} 
\newcommand{\EMinj}{{\epsilon^{M}_{inj}}} 

\newcommand{\Ekin}{\neutrenergysymb^K} 
\newcommand{\Emag}{\neutrenergysymb^M} 
 %
 %


\newcommand{\EkRF}{E^U} 

\newcommand{\FmagL}{b} 
\newcommand{\Fmag}{{\vec{\FmagL}}} 

\newcommand{\FmagLF}{{{}\hat{b}}} 
\newcommand{\FmagF}{\vec{\FmagLF}} 
\newcommand{\FmagAL}{a} 
\newcommand{\FmagA}{{\vec{\FmagAL}}} 
\newcommand{\FmagJ}{{\vec{j}}} 
\newcommand{\FmagAF}{\vec{\hat{a}}} 
\newcommand{\FmagAFk}{\FmagAF_{\vk}} 

\newcommand{\FvelL}{v} 
\newcommand{\Fvel}{{\vec{\FvelL}}} 

\newcommand{\FvelLF}{{{}\hat{v}}} 
\newcommand{\FvelF}{\vec{\FvelLF}} 


\newcommand{\FvelA}{{\vec{\FvelL_A}}} 


\newcommand{\cs}{c_s} 

\newcommand{\specw}{\zeta} 
\newcommand{\projk}{{\cal{P}^{\specw}}} 
\newcommand{\projkij}{{{\cal{P}}^{\specw}_{ij}}} 
\newcommand{\projkT}{\underline{\projk}} 


\newcommand{\kone}{\kappa} 

\newcommand{\rhowIIv}{\vec{w}} 
\newcommand{\rhowIIIv}{\vec{u}}


\newcommand{\IscHm}{{\cal{I}}_{\mhel}}

\newcommand{\rhoz}{\rho_0}





\newcommand{\expoalf}{\gamma}















\newcommand{\sm}{\text{-}}

\newcommand{\powerspec}{{\cal{P}}}
\newcommand{\powerspeccomp}{\powerspec^{\compexpo}}
\newcommand{\powerspecsol}{\powerspec^{\solexpo}}
\newcommand{\productspec}{{\cal{Q}}}

\newcommand{\press}{p} 

\newcommand{\tForceV}{\tilde{\vec{f}}_V}
\newcommand{\ForceV}{\vec{f}_V}

\newcommand{\tForceVijk}{\tilde{\vec{f}}_{V,i,j,k}}
\newcommand{\tForceM}{\tilde{\vec{f}}_M}
\newcommand{\ForceM}{\vec{f}_M}

\newcommand{\MVcA}{M5c}
\newcommand{\MVcB}{M5cB}

\newcommand{\VBalign}{\rho_{C}}
\newcommand{\mathI}{{\bf I}}

\renewcommand{\vec}[1]{\boldsymbol{#1}}

\newcommand{\Ampp}{B_{\vk}^+}
\newcommand{\phasep}{\theta_{\vk}^+}

\newcommand{\AmpV}{\neutrAmp_V}

\newcommand{\volavg}[1]{\langle #1 \rangle}

%% file: screl_arxiv.bbl
\begin{thebibliography}{10}

\bibitem{AMP06}
A.~Alexakis, P.~D. Mininni, and A.~Pouquet.
\newblock On the inverse cascade of magnetic helicity.
\newblock {\em The Astrophysical Journal 640}, pages 335--343, 2006.

\bibitem{ALF42}
H.~Alfv\'{e}n.
\newblock On the existence of electromagnetic-hydrodynamic waves.
\newblock {\em Arkiv f\"{o}r Matematik, Astronomi och Fysik 29B(2)}, pages
  1--7, 1942.

\bibitem{ALU13}
H.~Aluie.
\newblock Scale decomposition in compressible turbulence.
\newblock {\em Physica D 247}, pages 54--65, 2013.

\bibitem{BAP99}
D.~Balsara and A.~Pouquet.
\newblock The formation of large-scale structures in supersonic
  magnetohydrodynamic flows.
\newblock {\em Physics of Plasmas Vol. 6 No. 1}, pages 89--99, 1999.

\bibitem{BAL10}
D.~S. Balsara.
\newblock Multidimensional {HLLE} {R}iemann solver: Application to {Euler} and
  magnetohydrodynamic flows.
\newblock {\em Journal of Computational Physics 229}, pages 1970--1993, 2010.

\bibitem{BER97}
M.~A. Berger.
\newblock Magnetic helicity in a periodic domain.
\newblock {\em Journal of Geophysical Research 102 No. A2}, pages 2637--2644,
  1997.

\bibitem{BER99}
M.~A. Berger.
\newblock Introduction to magnetic helicity.
\newblock {\em Plasma Physics and Controlled Fusion 41}, pages B167--B175,
  1999.

\bibitem{BEM87}
J.~W. Bieber, P.~A. Evenson, and W.~H. Matthaeus.
\newblock Magnetic helicity of the {P}arker field.
\newblock {\em The Astrophysical Journal, 315}, pages 700--705, 1987.

\bibitem{BIS93}
D.~Biskamp.
\newblock {\em Nonlinear Magnetohydrodynamics}.
\newblock Cambridge University Press, 1993.

\bibitem{BRA01}
A.~Brandenburg.
\newblock The inverse cascade and nonlinear alpha-effect in simulations of
  isotropic helical hydromagnetic turbulence.
\newblock {\em The Astrophysical Journal 550}, pages 824--840, 2001.

\bibitem{BRL13}
A.~Brandenburg and A.~Lazarian.
\newblock Astrophysical hydromagnetic turbulence.
\newblock {\em Space Science Reviews 178}, pages 163--200, 2013.

\bibitem{BRS05}
A.~Brandenburg and K.~Subramanian.
\newblock Astrophysical magnetic fields and nonlinear dynamo theory.
\newblock {\em Physics Reports 417}, pages 1--209, 2005.

\bibitem{BUH14}
P.~Buchm\"{u}ller and C.~Helzel.
\newblock Improved accuracy of high-order {WENO} finite volume methods on
  cartesian grids.
\newblock {\em Journal of Scientific Computing 61}, pages 343--368, 2014.

\bibitem{CCE03}
Q.~Chen, S.~Chen, and G.~L. Eyink.
\newblock The joint cascade of energy and helicity in three-dimensional
  turbulence.
\newblock {\em Physics of Fluids 15}, pages 361--374, 2003.

\bibitem{CHB01}
M.~Christensson and M.~Hindmarsh.
\newblock Inverse cascade in decaying three-dimensional magnetohydrodynamic
  turbulence.
\newblock {\em Physical Review E 64, 056405}, 2001.

\bibitem{ELS04}
B.~G. Elmegreen and J.~Scalo.
\newblock Interstellar turbulence {I}: Observations.
\newblock {\em Annual Review of Astronomy and Astrophysics 42}, pages 211--273,
  2004.

\bibitem{ELS56}
W.~M. Els\"{a}sser.
\newblock Hydromagnetic dynamo theory.
\newblock {\em Reviews of Modern Physics 28, No. 2}, pages 135--163, 1956.

\bibitem{EMO00}
D.~F. Escande, P.~Martin, S.~Ortolani, A.~Buffa, P.~Franz, L.~Marrelli,
  E.~Martines, G.~Spizzo, S.~Cappello, A.~Murari, R.~Pasqualotto, and P.~Zanca.
\newblock Quasi-single-helicity reversed-field-pinch plasmas.
\newblock {\em Physical Review Letters 85, No. 8}, pages 1662--1665, 2000.

\bibitem{EVH88}
C.~R. Evans and J.~F. Hawley.
\newblock Simulation of magnetohydrodynamic flows: a constrained transport
  method.
\newblock {\em The Astrophysical Journal 332}, pages 659--677, 1988.

\bibitem{FED13}
C.~Federrath.
\newblock On the universality of supersonic turbulence.
\newblock {\em Monthly Notices of the Royal Astronomical Society 436}, pages
  1245--1257, 2013.

\bibitem{FRD10}
C.~Federrath, J.~Roman-Duval, R.~S. Klessen, W.~Schmidt, and M.-M. Mac~Low.
\newblock Comparing the statistics of interstellar turbulence in simulations
  and observations, solenoidal versus compressive turbulence forcing.
\newblock {\em Astronomy and Astrophysics 512, A81}, 2010.

\bibitem{FLE96}
R.~C. Fleck, Jr.
\newblock Scaling relations for the turbulent, non-self-gravitating, neutral
  component of the interstellar medium.
\newblock {\em The Astrophysical Journal 458}, pages 739--741, 1996.

\bibitem{FPL75}
U.~Frisch, A.~Pouquet, J.~L\'{e}orat, and A.~Mazure.
\newblock Possibility of an inverse cascade of magnetic helicity in
  magnetohydrodynamic turbulence.
\newblock {\em Journal of Fluid Mechanics 68 Part 4}, pages 769--778, 1975.

\bibitem{GMP11}
J.P. Graham, P.~D. Mininni, and A.~Pouquet.
\newblock High {R}eynolds number magnetohydrodynamic turbulence using a
  {L}agrangian model.
\newblock {\em Physical Review E 84, 016314}, 2011.

\bibitem{KET08}
D.~I. Ketcheson.
\newblock Highly efficient strong stability-preserving {Runge-Kutta} methods
  with low-storage implementations.
\newblock {\em Society for Industrial and Applied Mathematics Journal on
  Scientific Computing 30, No. 4}, pages 2113--2136, 2008.

\bibitem{KNP07}
A.~G. Kritsuk, M.~L. Norman, P.~Padoan, and R.~Wagner.
\newblock The statistics of supersonic isothermal turbulence.
\newblock {\em The Astrophysical Journal 665}, pages 416--431, 2007.

\bibitem{KUR96}
A.~Kumar and D.~M. Rust.
\newblock Interplanetary magnetic clouds, helicity conservation, and
  current-core flux-ropes.
\newblock {\em Journal of Geophysical Research 101}, pages 667--684, 1996.

\bibitem{LPR99}
D.~Levy, G.~Puppo, and G.~Russo.
\newblock Central {WENO} schemes for hyperbolic systems of conservation laws.
\newblock {\em Mathematical Modelling and Numerical Analysis 33, No. 3}, pages
  547--571, 1999.

\bibitem{LBM16}
M.~Linkmann, A.~Berera, M.~McKay, and J.~J\"{a}ger.
\newblock Helical mode interactions and spectral transfer processes in
  magnetohydrodynamic turbulence.
\newblock {\em Journal of Fluid Mechanics 791}, pages 61--96, 2016.

\bibitem{LSM17}
M.~Linkmann, G.~Sahoo, M.~McKay, A.~Berera, and L.~Biferale.
\newblock Effects of magnetic and kinetic helicities on the growth of magnetic
  fields in laminar and turbulent flows by helical fourier decomposition.
\newblock {\em The Astrophysical Journal 836:26}, 2017.

\bibitem{LOW94}
B.~C. Low.
\newblock Magnetohydrodynamic processes in the solar corona: Flares, coronal
  mass ejections, and magnetic helicity.
\newblock {\em Physics of Plasmas 1}, pages 1684--1690, 1994.

\bibitem{MCL99}
M.-M. Mac~Low.
\newblock The energy dissipation rate of supersonic, {MHD} turbulence in
  molecular clouds.
\newblock {\em The Astrophysical Journal 524}, pages 169--178, 1999.

\bibitem{MAL09}
S.~K. Malapaka.
\newblock {\em A Study of Magnetic Helicity in Decaying and Forced {3D-MHD}
  Turbulence}.
\newblock PhD thesis, Universit\"{a}t Bayreuth, 2009.

\bibitem{COC11}
P.~McCorquodale and P.~Colella.
\newblock A high-order finite-volume method for conservation laws on locally
  refined grids.
\newblock {\em Communications in Applied Mathematics and Computational Science
  6, No. 1}, pages 1--25, 2011.

\bibitem{MFP81}
M.~Meneguzzi, U.~Frisch, and A.~Pouquet.
\newblock Helical and nonhelical turbulent dynamos.
\newblock {\em Physical Review Letters 47, No. 15}, pages 1060--1064, 1981.

\bibitem{MIP09}
P.~D. Mininni and A.~Pouquet.
\newblock Finite dissipation and intermittency in magnetohydrodynamics.
\newblock {\em Physical Review E 80, 025401}, 2009.

\bibitem{MOF69}
H.~K. Moffatt.
\newblock The degree of knottedness of tangled vortex lines.
\newblock {\em Journal of Fluid Mechanics 35}, pages 117--129, 1969.

\bibitem{MMB12}
W.-C. M\"{u}ller, S.~K. Malapaka, and A.~Busse.
\newblock Inverse cascade of magnetic helicity in magnetohydrodynamic
  turbulence.
\newblock {\em Physical Review E 85, 015302}, 2012.

\bibitem{ORS70}
S.~A. Orszag.
\newblock Analytical theories of turbulence.
\newblock {\em Journal of Fluid Mechanics 41}, pages 363--386, 1970.

\bibitem{PFL76}
A.~Pouquet, U.~Frisch, and J.~L\'{e}orat.
\newblock Strong {MHD} helical turbulence and the nonlinear dynamo effect.
\newblock {\em Journal of Fluid Mechanics 77, part 2}, pages 321--354, 1976.

\bibitem{POP78}
A.~Pouquet and G.~S. Patterson.
\newblock Numerical simulation of helical magnetohydrodynamic turbulence.
\newblock {\em Journal of Fluid Mechanics 85, part 2}, pages 305--323, 1978.

\bibitem{RUS61}
V.~V. Rusanov.
\newblock The calculation of the interaction of non-stationary shock waves with
  barriers.
\newblock {\em Zhurnal Vychislitel'noi Matematiki i Matematicheskoi Fiziki,
  1:2}, pages 267--279, 1961.
\newblock English: USSR Computational Mathematics and Mathematical Physics,
  1:2, 304-320, 1962.

\bibitem{TEI20}
J.-M. Teissier.
\newblock {\em Magnetic helicity inverse transfer in isothermal supersonic
  magnetohydrodynamic turbulence}.
\newblock PhD thesis, Technische Universit\"{a}t Berlin, 2020.

\bibitem{VTH19}
P.~S. Verma, J.-M. Teissier, O.~Henze, and W.-C. M\"{u}ller.
\newblock Fourth order accurate finite volume {CWENO} scheme for astrophysical
  {MHD} problems.
\newblock {\em Monthly Notices of the Royal Astronomical Society 482}, pages
  416--437, 2019.

\bibitem{VIC01}
E.~T. Vishniac and J.~Cho.
\newblock Magnetic helicity conservation and astrophysical dynamos.
\newblock {\em The Astrophysical Journal 550}, pages 752--760, 2001.

\bibitem{WOL58a}
L.~Woltjer.
\newblock A theorem on force-free magnetic fields.
\newblock {\em Proceedings of the National Academy of Sciences 44, No. 6},
  pages 489--491, 1958.

\end{thebibliography}
